\documentclass[prd,onecolumn,notitlepage,eqsecnum,nofootinbib,floatfix]{revtex4-1}
\usepackage{amssymb}
\usepackage{amsmath}
\usepackage{amsfonts} 
\usepackage{bm}
\usepackage{graphicx}
\allowdisplaybreaks
\begin{document}
\title{Equilibrium and stability of thin spherical shells in Newtonian and relativistic gravity}  
\author{Philip LeMaitre and Eric Poisson}  
\affiliation{Department of Physics, University of Guelph, Guelph,
  Ontario, N1G 2W1, Canada} 
\date{September 13, 2019} 
\begin{abstract}
We consider thin spherical shells of matter in both Newtonian gravity and general relativity, and examine their equilibrium configurations and dynamical stability. Thin-shell models are admittedly a poor substitute for realistic stellar models. But the simplicity of the equations that govern their dynamics, compared with the much more complicated mechanics of a self-gravitating fluid, allows us to deliver, in a very direct and easy manner, powerful insights regarding their equilibria and stability. We explore, in particular, the link between the existence of a maximum mass along a sequence of equilibrium configurations and the onset of dynamical instability. Such a link is well-established in the case of fluid bodies in both Newtonian gravity and general relativity, but the demonstration of this link is both subtle and difficult. The proof is very simple, however, in the case of thin shells, and it is constructed with nothing more than straightforward algebra and a little calculus.   
\end{abstract} 
\maketitle

\section{Introduction} 
\label{sec:intro} 

Gravitational collapse to a black hole is assuredly one of the most cataclysmic events that can occur in the universe. It is also one of the most fascinating, and the subject is in high demand in courses devoted either to stellar structure and evolution, or to general relativity.

Stellar-mass black holes can form in a variety of scenarios. For example, a black hole forms when a supernova explosion leaves behind a core that is too massive to form a neutron star. As another example, a black hole forms when the members of a binary neutron-star system finally merge after a long stage of inspiral, giving rise to a powerful burst of gravitational waves and electromagnetic radiation \cite{GW170817:17, goldstein-etal:17, savchenko-etal:17}. For the purposes of this paper we have in mind the following situation. Consider a binary system that consists of a neutron star and a main-sequence star in a tight orbit. We imagine that the neutron star has been steadily accreting matter from its companion, its mass increasing in the process, and that it is just about to achieve the maximum mass allowed by its internal dynamics (governed by gravitation and nuclear physics). When the maximum mass is finally reached, the neutron star becomes dynamically unstable, and it promptly collapses to a black hole.

\begin{figure}
\includegraphics[width=0.7\linewidth]{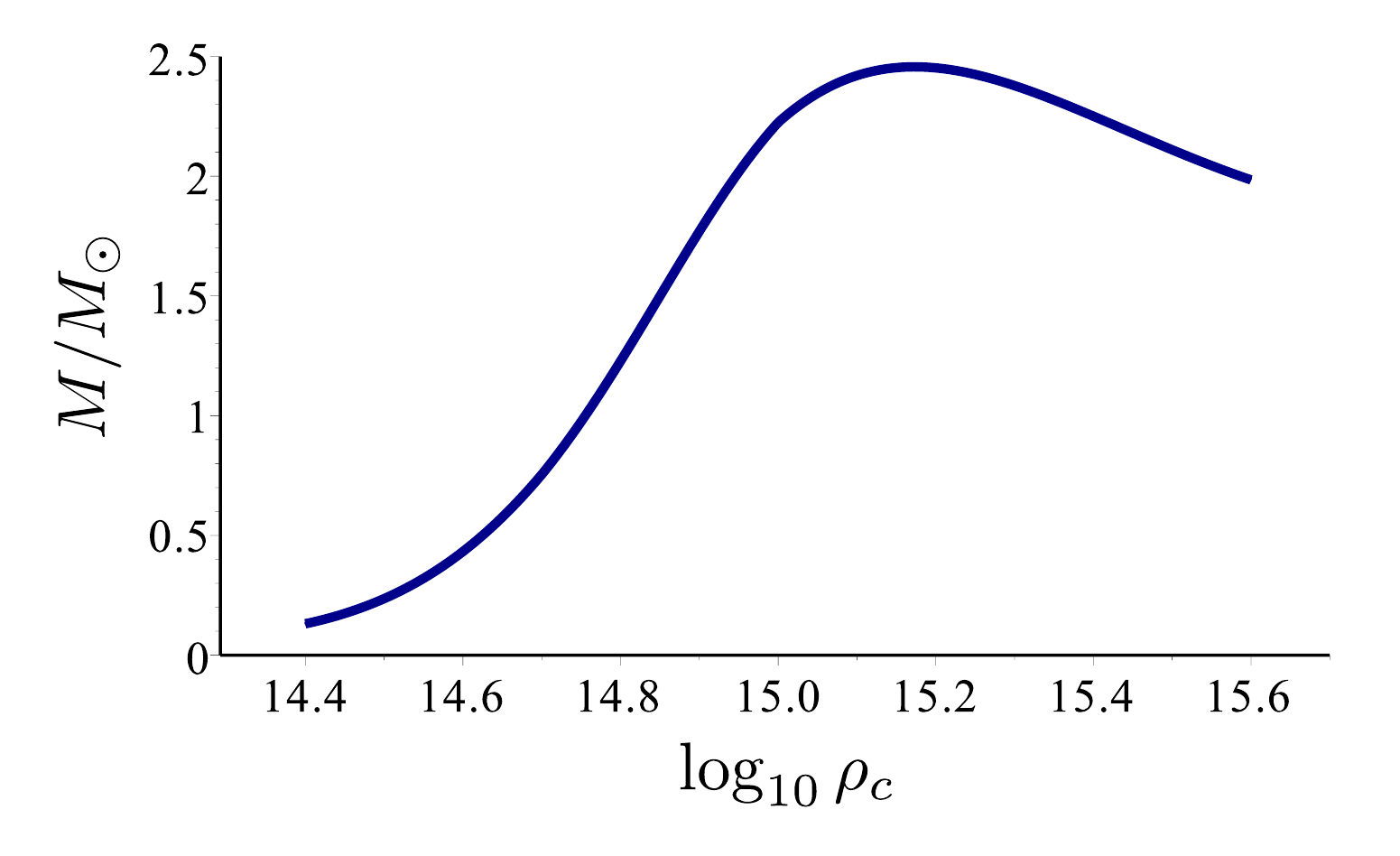}
\caption{Mass $M$ of a neutron star (in solar masses) as a function of $\log_{10}\rho_c$, the logarithm of the central mass density $\rho_c$ (in $\mbox{g}/\mbox{cm}^3$), for the MPA1 equation of state. The mass achieves a maximum of $M_{\rm max} \simeq 2.45\ M_\odot$ when $\log_{10}\rho_c \simeq 15.17$. For $M < M_{\rm max}$, the neutron-star configurations are dynamically stable. They are unstable for $M > M_{\rm max}$, and configurations beyond the maximum undergo gravitational collapse to a black-hole state.}  
\label{fig:fig1} 
\end{figure}

Our concern in this paper is the link between the maximum mass of a sequence of neutron-star equilibria and the onset of dynamical instability that leads to the formation of a black hole. A typical sequence is shown in Fig.~\ref{fig:fig1}, which plots the mass $M$ of a neutron star as a function of its central mass density $\rho_c := \rho(r=0)$. (A useful introduction to neutron stars can be found in Ref.~\cite{silbar-reddy:04}.) The sequence is calculated on the basis of the relativistic structure equations (see, for example, Chapter 24 of Ref.~\cite{hartle:03}) and the specific model known as MPA1 for the equation of state of nuclear matter \cite{muther-prakash-ainsworth:87}, one of a multitude of such models\footnote{The sequence is actually calculated using a piecewise-polytropic approximation to MPA1, as formulated in Ref.~\cite{read-etal:09b}.}. At low densities the mass is seen to increase with central density, as expected. But the mass achieves a maximum, $M_{\rm max} \simeq 2.45\  M_\odot$, when the central density reaches the critical value of $1.48 \times 10^{15}\ \mbox{g}/\mbox{cm}^3$, and it decreases beyond this point. In our scenario, the accreted matter is slowly deposited onto the neutron-star surface, the central density increases as a result, and the mass increases until $M_{\rm max}$ is achieved. It is at this point that the neutron star becomes unstable to gravitational collapse and implodes into a black hole. Simple descriptions of this collapse are provided in Refs.~\cite{gautreau-cohen:95, adler-etal:05}. 

The link between the maximum mass on an equilibrium sequence and the onset of dynamical instability is well understood by researchers on relativistic stellar structure and evolution, and most students of general relativity will be told of this connection (see, for example, Sec.~24.5 of Ref.~\cite{hartle:03}). But few students will see a proof of the statement that the onset of instability occurs at an extremum of the relation $M(\rho_c)$. Our aim with this paper is to offer some insights into this matter, in the context of a simple toy model involving a thin spherical shell of matter.

It is not surprising that the link between maximum mass and onset of instability is not presented in introductory courses on stellar structure and evolution, or introductory courses in general relativity. The reason is simple: the proof is long, and it requires machinery that is not accessible to students in introductory courses. For example, the student would have to master the theory of fluid perturbations in situations in which the stellar fluid is self-gravitating. In the context of Newtonian gravity, the proof occupies about 23 pages of the excellent textbook by Shapiro and Teukolsky \cite{shapiro-teukolsky:83}. In general relativity, the proof is unfolded in the 166 pages of a charming treatise by Harrison, Thorne, Wakano, and Wheeler \cite{harrison-thorne-wakano-wheeler:1965}. Working through these details requires enormous dedication.  

This situation is unsatisfying. The link between maximum mass and onset of instability is clearly extremely important in gravitational physics --- Newtonian and relativistic --- and the introductory student ought to be given more than just a mere statement of the fact. It should be possible to distill the details of the proof in a treatment that is accessible to this student. This is our aim in this paper. To capture the essence of the proof, we model a star as a thin spherical shell of matter. This is, of course, both artificial and unrealistic; stars are nothing like thin shells. But much is gained by this idealization: the fluid mechanics is radically simplified, the statics and dynamics of the star become fully accessible in both Newtonian and relativistic gravity, and the connection between maximum mass and onset of instability is brought out in a most vivid manner. To us, the insights provided by this simplicity outweigh, by a wide margin, the loss of physical realism brought upon by the radical idealization. The thin-shell model provides an accessible avenue to the introductory student to explore the link between maximum mass and onset of dynamical instability. 

We begin our developments in Sec.~\ref{sec:newtonian} with a discussion of the statics and dynamics of a thin spherical shell in Newtonian gravity. We derive the governing equations, explore equilibrium configurations under an assumed equation of state, calculate the change in mass in relation to a change of density along an equilibrium sequence, and identify the condition associated with a vanishing change (an extremum of the mass). We continue with an analysis of the stability of equilibrium configurations, and show that the condition for the onset of dynamical instability coincides with the condition for a mass extremum. In Sec.~\ref{sec:relativistic} we repeat these steps in the framework of general relativity. We conclude the paper in Sec.~\ref{sec:conclusion} with a recapitulation of our main results and a discussion of their significance. Some technical aspects are relegated to Appendices. In Appendix~\ref{sec:A} we resolve an ambiguity regarding the Newtonian gravitational force acting on a thin shell, and in Appendix~\ref{sec:B} we provide elements of derivation for the equations that govern the motion of a relativistic thin shell.    

Our considerations in this paper rely on the existence of an equation of state that relates the surface pressure $p$ of the thin shell to its surface density $\sigma$. We avoid making a specific choice of equation of state (except for the purposes of illustration in some examples), but we take it to have a barotropic form $p = p(\sigma)$. The key point is that the pressure is assumed to depend on density only; no other variables are allowed to enter the equation of state. For stars on the main sequence, this would be an untenable assumption; for these stars the equation of state depends on additional variables such as temperature and chemical composition. The barotropic form, however, is usually adequate to model a neutron star. For such stars the temperature is typically low compared with the Fermi temperature, and it can therefore be ignored in the equation of state. And because nuclear composition is a function of density, this dependence can be adequately captured by the barotropic form. Because our thin shells are proxies for neutron stars, it is appropriate to give them also an equation of state of the form $p = p(\sigma)$.

\section{Newtonian shell} 
\label{sec:newtonian} 

In this section we examine the statics and dynamics of a thin spherical shell of matter in Newtonian gravity. 

\subsection{Governing equations} 
\label{subsecN:governing} 

We consider a thin spherical shell of matter, in the limit in which the thickness is taken to be infinitesimal. The shell's radius is $R(t)$, its surface mass density is $\sigma(t)$, and its tangential surface pressure is $p(t)$. As discussed near the end of Sec.~\ref{sec:intro}, we assume that there exists an equation of state $p = p(\sigma)$ relating the pressure and density. The shell's mass is 
\begin{equation} 
m = 4\pi R^2 \sigma, 
\end{equation}
and this is a constant of the shell's motion. To derive the shell's equations of motion we take advantage of the spherical symmetry, which ensures that forces are directed either inward or outward, and which frees us from the need to perform a complete vectorial analysis of Newton's second law.    

The surface pressure $p$ pushes the shell outward. Its definition is provided by the statement that the work done by the shell as it changes its area by $dA$ is $dW = p\, dA$. Substituting $A = 4\pi R^2$ yields $dW = 8\pi p R\, dR$, and we recognize $8\pi p R$, the factor in front of the displacement $dR$, as the force acting on the shell. The pressure force per unit area is $2p/R$. 

The gravitational force pulls the shell inward. The force on an element of mass $dm$ situated just above the shell is $G m dm/R^2$, and integration yields $Gm^2/(2R^2)$ for the total force. The gravitational force per unit area is then $G\sigma m/(2R^2)$. We note that the gravitational force acting on the shell is actually ambiguous. The force acting on an element of mass situated just below the shell would be zero, and integration in this case would return a vanishing force. Which of the two results, if any, is valid? The proper way to obtain the force is to consider a shell that is thin but not infinitely thin, calculate the total force on the shell, and then observe that the result is independent of the mass distribution within the shell. Such a calculation is presented in Appendix~\ref{sec:A}, where it is confirmed that the gravitational force per unit area is indeed $G\sigma m/(2R^2)$.  

The net force acting on the shell is equal to its mass times its acceleration. Dividing by the area, we get $\sigma \ddot{R} = 2p/R - G\sigma m/(2R^2)$. Dividing by $\sigma = m/(4\pi R^2)$, we arrive at  
\begin{equation} 
\ddot{R} = -\frac{Gm}{2R^2} + \frac{8\pi p R}{m}.  
\label{Rdotdot_N} 
\end{equation} 
This is the shell's equation of motion.  

\subsection{Equilibrium configurations} 
\label{subsecN:equilib} 

An equilibrium configuration requires a precise balance between gravity and surface pressure, and Eq.~(\ref{Rdotdot_N}) reveals that the pressure must be given by 
\begin{equation} 
p = \frac{G m^2}{16\pi R^3}. 
\label{pequilib_N}
\end{equation} 
The shell's density is related to the mass by 
\begin{equation} 
\sigma = \frac{m}{4\pi R^2}. 
\end{equation} 
These equations can be solved to give $m$ and $R$ as functions of $p$ and $\sigma$. We find that 
\begin{equation} 
m = \frac{4}{\pi G^2} \frac{p^2}{\sigma^3}, \qquad 
R = \frac{1}{\pi G} \frac{p}{\sigma^2}. 
\label{mR_density} 
\end{equation} 
With $p$ related to $\sigma$ by an equation of state, we have that both $m$ and $R$ can be expressed as functions of the density only. 

It is useful to introduce a compactness parameter $C := G m/R$ to characterize an equilibrium configuration. This quantity is numerically equal to the square of the escape velocity from the shell's gravitational pull. According to Eqs.~(\ref{mR_density}), it is given by 
\begin{equation} 
C = \frac{4p}{\sigma}, 
\label{C_N} 
\end{equation} 
and it provides a measure of the pressure-to-density ratio.  

The results obtained thus far are quite general, and they do not require a specific choice of equation of state. Choices, however, must be made to provide illustrations of these results. We shall consider two simple examples of barotropic equations of state.

Our first example is the polytropic form $p = K \sigma^\gamma$, where $K$ and $\gamma$ are constants. For this we find that Eqs.~(\ref{mR_density}) become 
\begin{equation} 
m = \frac{4K^2}{\pi G^2} \sigma^{2\gamma-3}, \qquad 
R = \frac{K}{\pi G} \sigma^{\gamma-2}. 
\end{equation} 
We see that $m$ increases with $\sigma$ when $\gamma > 3/2$, and that it decreases when $\gamma < 3/2$.
We also see that $R$ decreases with increasing $\sigma$ when $\gamma < 2$, and that it increases when $\gamma > 2$. The mass and radius can also be expressed in terms of the compactness parameter. According to Eq.~(\ref{C_N}), this is related to the density by $C = 4K \sigma^{\gamma-1}$, and we find that 
\begin{equation} 
m = \frac{(4K)^{\frac{1}{\gamma-1}}}{4\pi G^2} C^{\frac{2\gamma-3}{\gamma-1}},  \qquad 
R = \frac{(4K)^{\frac{1}{\gamma-1}}}{4\pi G} C^{\frac{\gamma-2}{\gamma-1}}. 
\label{mR_C} 
\end{equation}  
Because the mass is a monotonic function of density, we observe that the polytropic equation of state does not produce a turning point in the function $m(\sigma)$, of the sort seen in Fig.~\ref{fig:fig1} for a realistic model of neutron star. Instead, the maximum mass is achieved either when $\sigma = \infty$ ($\gamma > 3/2$) or when $\sigma = 0$ ($\gamma < 3/2$); and this maximum mass is infinite. This model, therefore, does not quite capture what we had in mind. 

\begin{figure} 
\includegraphics[width=0.7\linewidth]{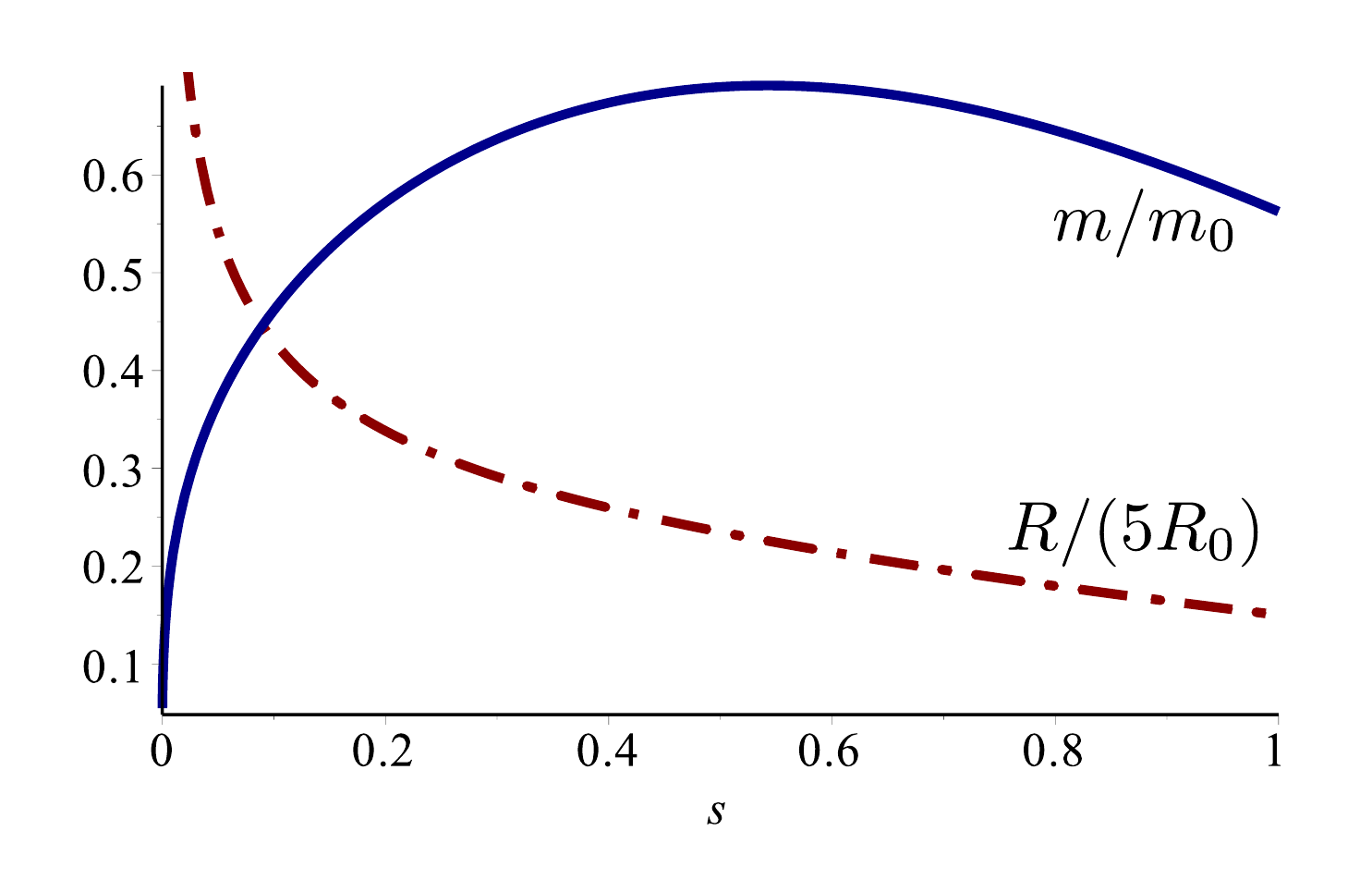}
\caption{Radius and mass as functions of $s := \sigma/\sigma_0$ for the sinusoidal equation of state. The dotted-dashed red curve shows $R/(5R_0)$ and the solid blue curve shows $m/m_0$.} 
\label{fig:fig2} 
\end{figure} 

Another example of an equation of state --- a better one for our purposes --- is 
\begin{equation} 
p = p_0 \bigl[ \sin(\sigma/\sigma_0) \bigr]^{5/3}, 
\label{sinEOS}
\end{equation} 
where $p_0$ and $\sigma_0$ are constants; it is understood that $\sigma/\sigma_0 < \pi/2$, to ensure that $p$ strictly increases with $\sigma$. For this we have that 
\begin{equation} 
m = m_0 \frac{(\sin s)^{10/3}}{s^3}, \qquad 
R = R_0 \frac{(\sin s)^{5/3}}{s^2}, 
\end{equation} 
where $s := \sigma/\sigma_0$, $m_0 := 4 p_0^2/(\pi G^2 \sigma_0^3)$, and $R_0 := p_0/(\pi G \sigma_0^2)$. Plots of these functions are displayed in Fig.~\ref{fig:fig2}. We find that $m$ increases from $m = 0$ when $s = 0$ to a maximum value $m_{\rm max} \simeq 0.6914\, m_0$ when $s \simeq 0.5423$, and that it decreases thereafter; the value of the mass at $s = 1$ is $m \simeq 0.5625\, m_0$. On the other hand, $R$ is seen to be monotonically decreasing from $R = \infty$ when $s = 0$ to $R \simeq 0.7500\, R_0$ when $s = 1$. For this equation of state we have that $m(\sigma)$ behaves in a way that is qualitatively similar to the neutron-star mass of Fig.~\ref{fig:fig1}; this model produces a turning point, a finite maximum mass, and it fully captures what we had in mind.  

It is instructive to define an effective polytropic exponent $\gamma(\sigma)$ for the equation of state of Eq.~(\ref{sinEOS}). The definition is provided by the relation $dp/p = \gamma\, d\sigma/\sigma$, and a simple computation returns 
\begin{equation} 
\gamma = \frac{5}{3} \frac{s \cos s}{\sin s}. 
\end{equation} 
The exponent is monotonically decreasing with increasing $\sigma$. Its value is $\gamma = 5/3 \simeq 1.667$ when $s=0$, and it decreases to $\gamma \simeq 1.0702$ when $s = 1$. Its value when $m$ reaches its maximum, at $s \simeq 0.5423$, is $\gamma \simeq 1.500$. The equation of state of Eq.~(\ref{sinEOS}) was designed specifically so that $\gamma$ would be a decreasing function of $s$. 

In the case of the polytropic equation of state, we have seen that $m$ increases with $\sigma$ when $\gamma > 3/2$, and that it decreases when $\gamma < 3/2$; in this context the value of $\gamma$ is constant on the entire equilibrium sequence. In the case of the sinusoidal equation of state, we have an effective polytropic exponent $\gamma$ that varies with density, but it is still true that $m$ increases with density while $\gamma > 3/2$, and that it decreases when $\gamma$ drops below $3/2$. As we shall see below, this observation is not limited to these specific examples; it is true of any equilibrium sequence that $m$ increases while $\gamma(\sigma) > 3/2$, reaches a maximum at $\gamma(\sigma) = 3/2$, and decreases beyond $\gamma(\sigma) = 3/2$.   

\subsection{Changes in mass and radius along an equilibrium sequence} 
\label{subsecN:sequence} 

We return to Eqs.~(\ref{mR_density}), which describe an equilibrium configuration, and consider a sequence of such configurations parametrized by $\sigma$, the shell's mass density. We continue to assume that the surface pressure $p$ is related to $\sigma$ by a barotropic equation of state, but we make no specific choice for this equation of state.  

As we vary $\sigma$ and $p$ along the equilibrium sequence, we find that $m$ and $R$ change according to 
\begin{equation} 
dm = \frac{4}{\pi G^2} \biggl( \frac{2p}{\sigma^3}\, dp - \frac{3p^2}{\sigma^4}\, d\sigma \biggr), \qquad 
dR = \frac{1}{\pi G} \biggl( \frac{1}{\sigma^2}\, dp - \frac{2p}{\sigma^3}\, d\sigma \biggr). 
\end{equation} 
The change in $p$ is related to the change in $\sigma$ by the equation of state, and $dp$ is therefore not independent of $d\sigma$. We express the relation in terms of the effective polytropic exponent $\gamma(\sigma)$ introduced previously, which is defined by $dp/p = \gamma\, d\sigma/\sigma$ for any equation of state. With this we have that 
\begin{equation} 
dm = \frac{4}{\pi G^2} \frac{p^2}{\sigma^4} ( 2\gamma - 3 )\, d\sigma, \qquad 
dR = \frac{1}{\pi} \frac{p}{\sigma^3} ( \gamma - 2 )\, d\sigma, 
\end{equation} 
or 
\begin{equation} 
dm = 4\pi R^2 (2\gamma -3)\, d\sigma, \qquad 
dR = \frac{4\pi R^3}{m} ( \gamma - 2 )\, d\sigma 
\end{equation} 
after making use of Eqs.~(\ref{mR_density}). The first equation reveals that for any equation of state, $m$ increases with density when $\gamma(\sigma) > 3/2$, and that it decreases when $\gamma(\sigma) < 3/2$. When $\gamma$ is a monotonically decreasing function of the density, we have that $m$ increases while $\gamma > 3/2$, that it reaches a maximum when $\gamma = 3/2$, and that it decreases beyond this point. This is the behavior observed previously with the sinusoidal equation of state, but we see now that the relation between $dm/d\sigma$ and $\gamma - 3/2$ is completely general. The second equation reveals that $R$ decreases with increasing density when $\gamma(\sigma) < 2$, and that it increases when $\gamma(\sigma) > 2$; when $\gamma = 2$ the radius is independent of the density.   

\subsection{Stability} 
\label{subsecN:stability} 

Whether an equilibrium configuration is dynamically stable can be decided by perturbing the equilibrium and using 
Eq.~(\ref{Rdotdot_N}) to determine the evolution of the perturbation. Denoting the equilibrium values with a zero-subscript, we have that 
\begin{equation} 
p_0 = \frac{G m^2}{16\pi R_0^3}, \qquad 
\sigma_0 = \frac{m}{4\pi R_0^2}. 
\end{equation} 
The equilibrium is perturbed by letting $R$, $\sigma$, and $p$ deviate from their equilibrium values; $m$ is a constant of the shell's motion and it therefore stays equal to its equilibrium value. We write $R = R_0 + \delta R$, $\sigma = \sigma_0 + \delta \sigma$, and $p = p_0 + \delta p$, with the deviations $\delta R$, $\delta \sigma$, and $\delta p$ depending on time. These are not all independent. Because $m$ stays constant, we have that $\delta\sigma = -(m/2\pi R_0^3)\, \delta R$. And the equation of state implies that $\delta p$ can be related to $\delta \sigma$; we write the relation as $\delta p/p_0 = \Gamma\, \delta\sigma/\sigma_0$, in terms of an adiabatic exponent $\Gamma$ that depends on $\sigma_0$. In principle, $\Gamma$ should be distinguished from $\gamma$ introduced previously in the context of equilibrium sequences. The previous exponent, defined by $dp/p = \gamma\, d\sigma/\sigma$, characterizes a variation of the pressure and density along a sequence of equilibrium configurations. The exponent $\Gamma$, on the other hand, characterizes a dynamical variation of the pressure and density away from the equilibrium sequence. However, because the equation of state is assumed to be of the barotropic form $p = p(\sigma)$, the two exponents are actually equal when evaluated at the same reference density. 

Incorporating these ingredients in Eq.~(\ref{Rdotdot_N}) and linearizing with respect to the primary variable $\delta R$, we arrive at the perturbation equation 
\begin{equation} 
\delta \ddot{R} = -\frac{Gm}{2 R_0^3} (2\Gamma-3)\, \delta R. 
\end{equation} 
This is recognized as the equation of a simple harmonic oscillator, with $Gm(2\Gamma-3)/(2R_0^3)$ identified as the square of the oscillator's natural frequency. The equation implies that the perturbation will be oscillating when $\Gamma > 3/2$ (which gives rise to a real frequency), but growing exponentially when $\Gamma < 3/2$ (an imaginary frequency). The equilibrium is therefore dynamically stable when $\Gamma > 3/2$, and unstable when $\Gamma < 3/2$.  

In view of the equality $\Gamma = \gamma$ for any equation of state of the form $p = p(\sigma)$, we see that the condition for dynamical stability is identical to the condition for $dm_0/d\sigma_0 > 0$. In other words,  the equilibrium sequence is dynamically stable so long as the mass $m_0$ increases with density, and the onset of instability occurs when $m_0$ reaches a maximum. This is the punchline of this section of the paper.  

This conclusion rests heavily on the assumed barotropic form for the equation of state. For a more general equation of state like $p = p(\sigma,s)$, the specific entropy $s$ would vary along a sequence of equilibrium configurations, but it would be constant for a dynamical perturbation that does not provoke a flow of heat. In this broader context, $\gamma$ would be defined in terms of changes $dp$ and $d\sigma$ that are accompanied by a change in specific entropy, while $\Gamma$ would be defined in terms of adiabatic changes. The exponents would not be equal, and the onset of dynamical instability would no longer coincide with $dm_0/d\sigma_0 = 0$.

\section{Relativistic shell} 
\label{sec:relativistic} 

In this section we consider the statics and dynamics of a thin spherical shell of matter in general relativity. The mechanics of an electrically charged shell was thoroughly explored by Chase in Ref.~\cite{chase:70}. The situation considered here, in which the shell is uncharged, is a special case of Chase's analysis. Throughout we use relativistic units and set $G = c = 1$. Mass and time are thus measured in units of length, and surface density and pressure are measured in units of $1/\mbox{length}$.   

\subsection{Governing equations} 
\label{subsecR:governing} 

The equations that govern the behavior of a thin spherical shell in general relativity are similar to the Newtonian equations, but there are important conceptual differences in addition to the expected relativistic corrections. A first conceptual difference concerns the measurement of time. While there is a unique and universal notion of time in Newtonian mechanics, the same is not true in relativity, and one must therefore make a specific choice of time. In this section we shall adopt proper time $\tau$ as a time standard to describe the motion of the shell. By this we mean the following: we imagine an observer attached to the shell, moving along with it, and measuring time according to a standard clock that she carries with her; proper time $\tau$ is the time recorded by this clock.

A second conceptual difference concerns the notion of mass. There are, in fact, two notions of shell mass in the relativistic setting. The first is the same as in the Newtonian discussion: the shell's {\it material mass} $m$ is the product of the surface density $\sigma$ and the shell's area $A$. Defining a radius $R$ such that $A = 4\pi R^2$, we have that the material mass is given by 
\begin{equation} 
m = 4\pi R^2 \sigma.
\label{m_def} 
\end{equation}
When the shell moves, its radius $R$ and density $\sigma$ vary with proper time $\tau$, and $m$ also changes with time. {\it Unlike the Newtonian mass, the material mass is not a constant of the shell's motion.} The reason for this important difference is that in relativity, mass is also a measure of energy, and the shell's internal energy changes as the pressure does work on the shell. We have that $dm/d\tau = -p\, dA/d\tau$, or  
\begin{equation} 
\dot{m} = -8\pi p R\, \dot{R},  
\label{firstlaw} 
\end{equation} 
in which an overdot indicates differentiation with respect to $\tau$.

The second notion of shell mass is the {\it gravitational mass} $M$, which {\it is} a constant of the shell's motion. This is defined as follows. The shell at radius $r = R(\tau)$ can be thought to partition spacetime into two regions, the interior and exterior of the shell. The shell's interior is empty of matter, and there is no gravitational field inside the shell; this portion of spacetime is therefore flat, and described by the Minkowski metric of special relativity. The shell's exterior is also empty, but the gravitational field is no longer zero; this portion of spacetime is described by the Schwarzschild metric of general relativity, and the metric comes with a mass parameter $M$ that we identify with the shell's gravitational mass. This mass parameter is a measure of the total energy contained in the spacetime, which includes the shell's rest-mass energy, kinetic energy, and gravitational potential energy. The mathematical expression of this is    
\begin{equation} 
M = m \sqrt{1 + \dot{R}^2} - \frac{m^2}{2R}. 
\label{M_eq} 
\end{equation} 
As the shell moves, both $m$ and $R$ change with proper time $\tau$, but the total energy measured by $M$ stays constant. In the nonrelativistic limit, $\dot{R}^2 \equiv \dot{R}^2/c^2$ and $m/R \equiv Gm/(c^2 R)$ are very small, and neglecting these terms, we have that $M \to m$; there is only one notion of shell mass in Newtonian gravity. 

If we differentiate Eq.~(\ref{M_eq}) with respect to $\tau$, make use of Eq.~(\ref{firstlaw}) to eliminate $\dot{m}$, and then solve for $\ddot{R}$, we find that 
\begin{equation} 
\ddot{R} = -\sqrt{1+\dot{R}^2} \biggl[ \frac{m}{2R^2} 
- \frac{8\pi p R}{m} \Bigl( \sqrt{1+\dot{R}^2} - m/R \Bigr) \biggr]. 
\label{Rdotdot} 
\end{equation} 
This is the shell's equation of motion in general relativity. This equation replaces the Newtonian version of Eq.~(\ref{Rdotdot_N}), and we see that it incorporates a number of relativistic corrections. It is easy to see that we recover Eq.~(\ref{Rdotdot_N}) in the nonrelativistic limit. 

Equations (\ref{m_def}), (\ref{firstlaw}), (\ref{M_eq}), and (\ref{Rdotdot}) provide a foundation for a complete description of a thin spherical shell in general relativity. In Appendix~\ref{sec:B} we outline a derivation of these equations. 

\subsection{Equilibrium configurations} 
\label{subsecR:equilibria} 

We may now examine the equilibrium of a thin spherical shell in general relativity. We take all quantities to be independent of proper time $\tau$, and set $\dot{R} = 0 = \ddot{R}$ in the preceding equations. We observe that Eq.~(\ref{m_def}) stays unchanged, that Eq.~(\ref{firstlaw}) reduces to $0=0$, that Eq.~(\ref{M_eq}) becomes
\begin{equation} 
M = m \biggl( 1 - \frac{m}{2R} \biggr), 
\label{eq3}
\end{equation}
and that Eq.~(\ref{Rdotdot}) produces
\begin{equation}
p = \frac{m^2}{16\pi R^3(1-m/R)},  
\label{eq4} 
\end{equation}
an expression for the pressure that generalizes Eq.~(\ref{pequilib_N}). 

It is useful to introduce a compactness parameter 
\begin{equation}
C := \frac{m}{R} \equiv \frac{Gm}{Rc^2} 
\label{Cdef} 
\end{equation} 
to characterize an equilibrium configuration. Using Eq.~(\ref{eq3}) to solve for $m$ in terms of $M$, we find that $C$ can also be expressed as 
\begin{equation} 
C = 1 - \sqrt{1-2M/R}.   
\end{equation} 
When $M/R \equiv GM/(c^2 R)$ is small, $C$ is approximately equal to $M/R$ (and still exactly equal to $m/R$). We see that $C$ is necessarily smaller than $1$, since $R > 2M \equiv 2GM/c^2$. 

The foregoing equations can be manipulated to express $R$, $m$, and $M$ in terms of $\sigma$ and $p$. We first insert Eq.~(\ref{m_def}) within Eq.~(\ref{eq4}) and solve for $R$, then substitute this result back into Eq.~(\ref{m_def}), and finally make use of Eq.~(\ref{eq3}). We arrive at 
\begin{subequations} 
\label{RmM_sigmap} 
\begin{align} 
R &= \frac{p}{\pi \sigma^2(1 + 4p/\sigma)}, \\ 
m &= \frac{4p^2}{\pi \sigma^3(1 + 4p/\sigma)^2}\,  \\ 
M &= \frac{4p^2(1+2p/\sigma)}{\pi \sigma^3(1 + 4p/\sigma)^3}. 
\end{align} 
\end{subequations} 
From Eq.~(\ref{Cdef}) we also get 
\begin{equation} 
C = \frac{4p/\sigma}{1 + 4p/\sigma}, 
\label{C_sigmap} 
\end{equation} 
and note that $C$ depends on density and pressure through the ratio $p/\sigma$ only. Because $p$ can be related to $\sigma$ by the equation of state $p = p(\sigma)$, we have that the radius and mass parameters of an equilibrium configuration can be expressed in terms of the density only.  Equations (\ref{RmM_sigmap}) and (\ref{C_sigmap}) should be compared with Eqs.~(\ref{mR_density}) and (\ref{C_N}), their Newtonian limit. 

Equation (\ref{C_sigmap}) can be solved for $p/\sigma$, and we get 
\begin{equation} 
p/\sigma = \frac{C}{4(1-C)}. 
\label{p_over_sigma} 
\end{equation} 
This relation and the equation of state allow us to express the density as a function of the compactness parameter: $\sigma = \sigma(C)$. Equations (\ref{RmM_sigmap}) can then be rewritten in the useful alternative forms
\begin{subequations} 
\label{RmM_C} 
\begin{align} 
R &= \frac{C}{4\pi \sigma(C)}, \\
m &= C R = \frac{C^2}{4\pi \sigma(C)}, \\
M &= m(1 - \tfrac{1}{2} C) = \frac{C^2 (1-\tfrac{1}{2} C)}{4\pi \sigma(C)}. 
\end{align} 
\end{subequations} 

\begin{figure} 
\includegraphics[width=0.7\linewidth]{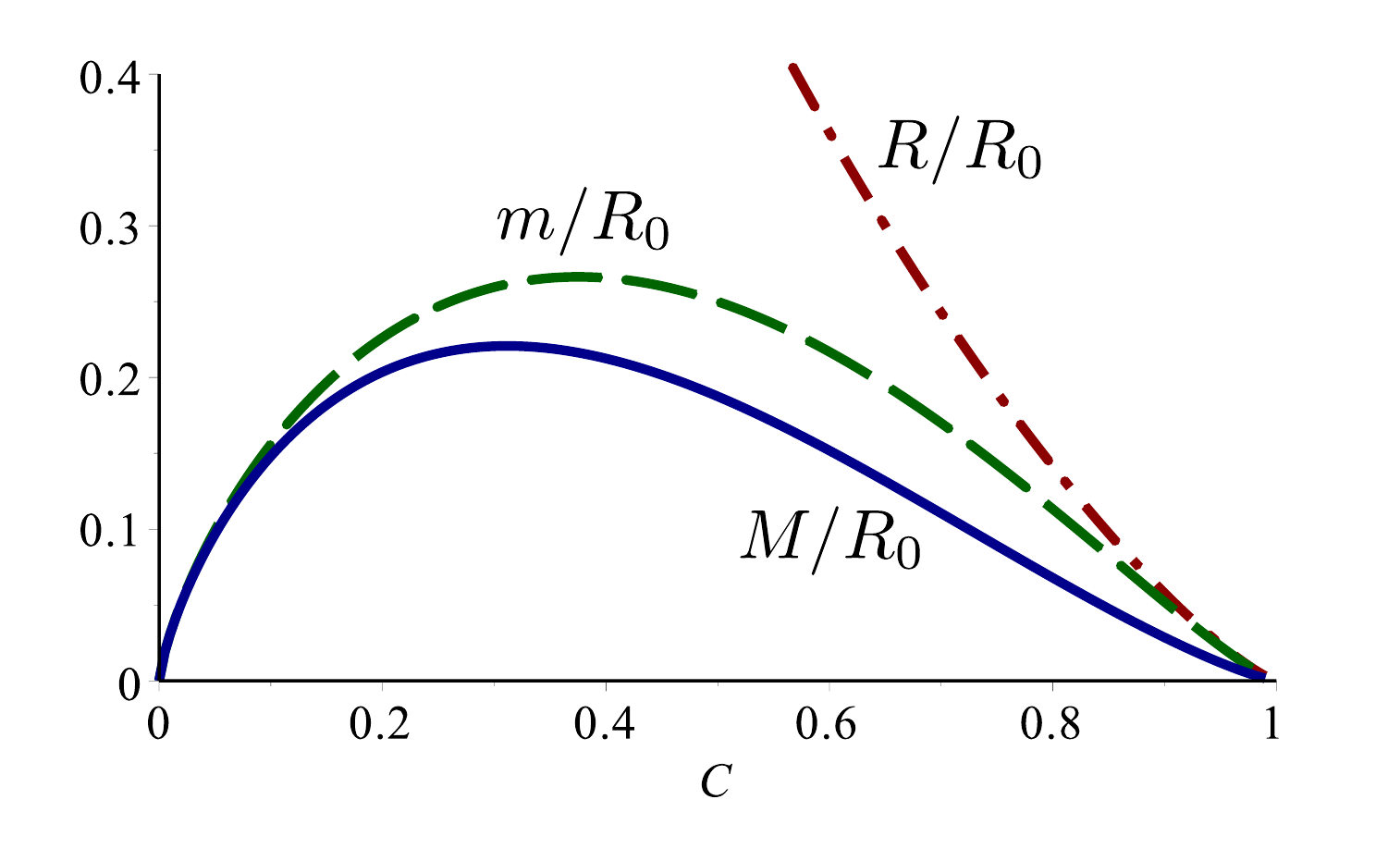}
\caption{Radius and mass parameters as functions of compactness for a polytropic equation of state with $\gamma = 1.8$. The dotted-dashed red curve shows $R/R_0$, the dashed green curve shows $m/R_0$, and the solid blue curve shows $M/R_0$. The unit of length $R_0$ is defined in Eq.~(\ref{R0_def}).}   
\label{fig:fig3} 
\end{figure} 

Our results thus far are quite general, and are valid for any barotropic equation of state. To formulate concrete examples of equilibrium configurations, we now suppose that the equation of state is specifically of the polytropic form $p = K \sigma^\gamma$, where $K$ and $\gamma$ are constants. In this case Eq.~(\ref{p_over_sigma})
gives $4 K \sigma^{\gamma-1} = C/(1-C)$, or 
\begin{equation} 
\sigma = \frac{1}{(4K)^{\frac{1}{\gamma-1}}} 
\frac{C^{\frac{1}{\gamma-1}}}{(1-C)^{\frac{1}{\gamma-1}}}. 
\end{equation} 
With this Eqs.~(\ref{RmM_C}) become
\begin{subequations} 
\begin{align} 
R &= R_0\, C^{\frac{\gamma-2}{\gamma-1}} (1-C)^{\frac{1}{\gamma-1}}, \\ 
m &= R_0\, C^{\frac{2\gamma-3}{\gamma-1}} (1-C)^{\frac{1}{\gamma-1}}, \\ 
M &= R_0\, C^{\frac{2\gamma-3}{\gamma-1}} (1-C)^{\frac{1}{\gamma-1}} (1 - \tfrac{1}{2} C),  
\end{align}
\end{subequations}  
where 
\begin{equation} 
R_0 := \frac{(4K)^{\frac{1}{\gamma-1}}}{4\pi} 
\label{R0_def} 
\end{equation} 
provides a unit of length and mass. We have that the radius and mass parameters can be written explicitly as functions of the compactness. In Fig.~\ref{fig:fig3} we show plots of $R/R_0$, $m/R_0$, and $M/R_0$ as functions of $C$ for $\gamma = 1.8$. We observe that $M$ achieves a maximum of $M \simeq 0.2208\, R_0$ at $C \simeq 0.3115$. Because $C$ increases monotonically with $\sigma$ (when $\gamma > 1$), we see that $M(\sigma)$ behaves qualitatively as the neutron-star mass of Fig.~\ref{fig:fig1}. We recall that in the Newtonian theory (see Sec.~\ref{subsecN:equilib}), the polytropic equation of state did not produce a mass function with a turning point; it does so in the relativistic theory.  

\subsection{Changes in mass and radius along an equilibrium sequence} 
\label{subsecR:sequence} 

We consider a sequence of equilibrium configurations corresponding to a given equation of state $p = p(\sigma)$. We note first that the equation of state implies that a change in $\sigma$ along the sequence must come with a specific change in $p$. As usual we express this relationship as $dp/p = \gamma\, d\sigma/\sigma$, in terms of the effective polytropic exponent $\gamma(\sigma)$.  

The sequence can be parametrized by the surface density $\sigma$, and the equations that describe this sequence are listed in Eqs.~(\ref{RmM_sigmap}) and (\ref{C_sigmap}). Alternatively, the sequence can be parametrized by the compactness $C$, and Eqs.~(\ref{RmM_C}) are the relevant equations for this parametrization; the density is then implicitly related to $C$ by Eq.~(\ref{p_over_sigma}). 

It is important to identify the conditions that ensure that $C$ and $\sigma$ are monotonic functions of each other. First, we note that according to Eq.~(\ref{C_sigmap}), $C$ is a monotonically increasing function of $x := p/\sigma$, because $dC/dx = 4/(1+4x)^2 > 0$. Second, we have that $dx/d\sigma = (\gamma-1) p/\sigma^2$, so that $x$ is monotonically increasing with $\sigma$ when $\gamma(\sigma) > 1$. Putting these results together, we have that 
\begin{equation}
dC = \frac{4(\gamma-1) p}{\sigma^2(1 + 4p/\sigma)^2}\, d\sigma, 
\label{dC_ds} 
\end{equation} 
and conclude that $C$ is monotonically increasing with $\sigma$ whenever $\gamma(\sigma) > 1$. Henceforth we shall restrict our attention to equations of state that satisfy this condition.     

Differentiation of Eqs.~(\ref{RmM_C}) and use of Eq.~(\ref{dC_ds}) yields 
\begin{subequations} 
\label{changes_sequence} 
\begin{align} 
dR &= -\frac{4\pi R^2}{C} \bigl[ 2 - \gamma + (\gamma - 1)C \bigr]\, d\sigma, \\ 
dm &= 4\pi R^2 \bigl[ 2\gamma - 3 - 2(\gamma - 1) C \bigr]\, d\sigma, \\ 
dM &= 2\pi R^2 \bigl[ 2(2\gamma-3) - (7\gamma-8) C + 3(\gamma-1) C^2 \bigr]\, d\sigma. 
\end{align} 
\end{subequations} 
These are the changes in $R$, $m$, and $M$ that accompany a change in $\sigma$ along the sequence of equilibrium configurations. 

\begin{figure} 
\includegraphics[width=0.7\linewidth]{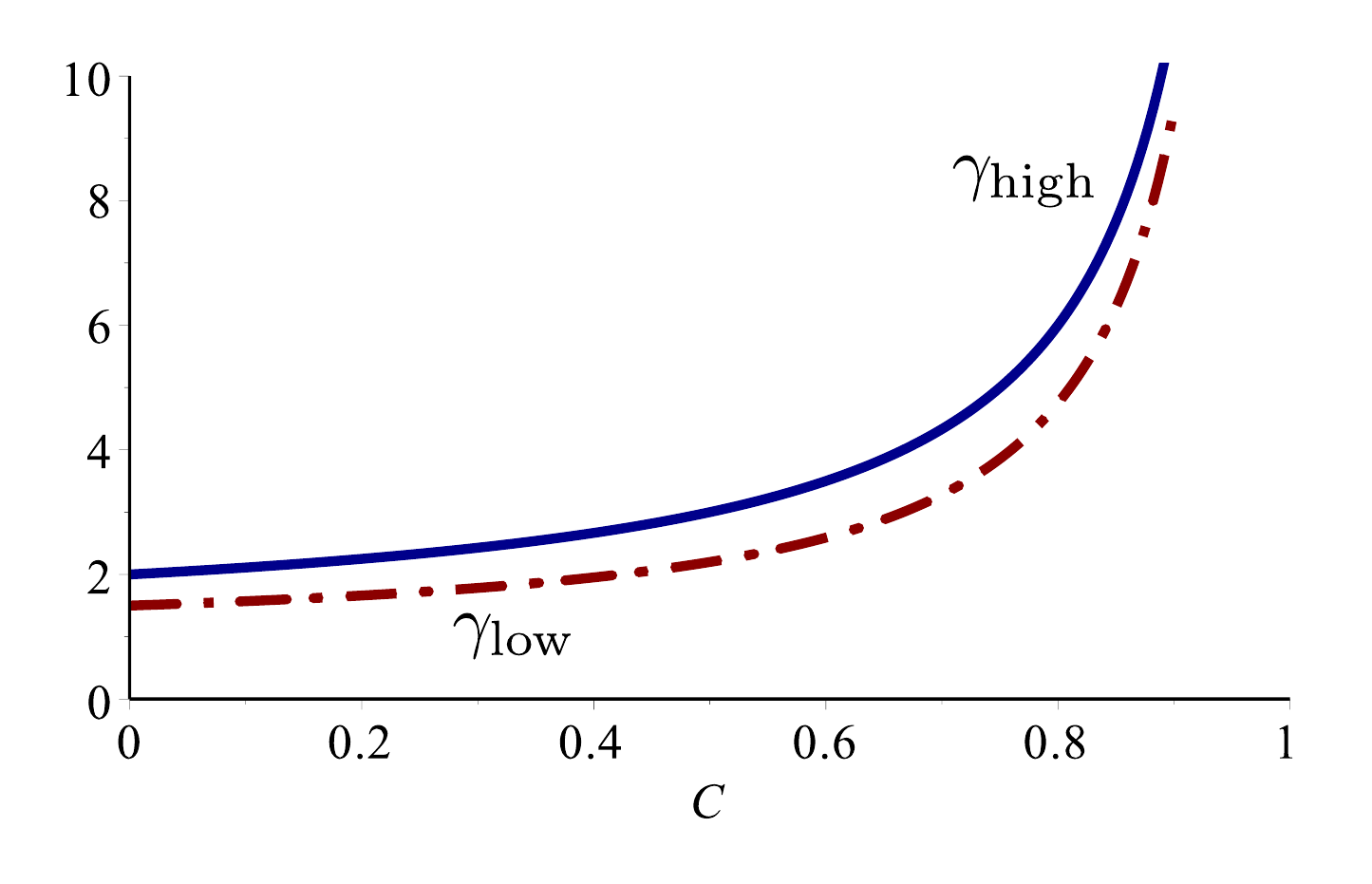}
\caption{The graph of $\gamma_{\rm high}$ as a function of $C$ is displayed in solid blue. The graph of $\gamma_{\rm low}$ as a function of $C$ is displayed in dotted-dashed red. When $\gamma(C) > \gamma_{\rm low}$, the mass $M$ increases with density $\sigma$. When $\gamma(C) < \gamma_{\rm low}$, the mass $M$ decreases with increasing density.} 
\label{fig:fig4} 
\end{figure}

It is reasonable to expect that $R$ should be decreasing as $\sigma$ increases. This requirement implies the following condition on the effective polytropic exponent: 
\begin{equation} 
\frac{dR}{d\sigma} < 0 \qquad \Longrightarrow \qquad 
\gamma(C) < \gamma_{\rm high} := \frac{2-C}{1-C}. 
\label{Gamma_hi} 
\end{equation} 
The function $\gamma_{\rm high}$ is plotted in Fig.~\ref{fig:fig4}. When $C$ is small, we have that $\gamma_{\rm high} \simeq 2$. 

The requirement that $M$ increases with increasing $\sigma$ returns the condition 
\begin{equation} 
\frac{dM}{d\sigma} > 0 \qquad \Longrightarrow \qquad 
\gamma(C) > \gamma_{\rm low} :=\frac{6-8C+3C^2}{(1-C)(4-3C)}. 
\label{Gamma_lo} 
\end{equation}
The function $\gamma_{\rm low}$ is plotted in Fig.~\ref{fig:fig4}. When $C$ is small, we have that $\gamma_{\rm low} = 3/2 + 5C/8 + O(C^2)$. 

We could examine, in a similar way, the condition associated with $dm/d\sigma > 0$, but we shall not go into these details here. In the next subsection we will reveal the connection between a point along the sequence at which $dM/d\sigma = 0$ and the onset of dynamical instability for the sequence. There appears to be no particular significance to a point at which $dm/d\sigma = 0$.    

\subsection{Dynamical stability} 
\label{subsecR:stability} 

Whether an equilibrium configuration is dynamically stable can be decided on the basis of a perturbative analysis in which the shell is taken away from equilibrium. The equilibrium is stable when the perturbed shell oscillates about the equilibrium configuration; it is unstable when the motion grows unboundedly. The governing equations were listed back in Sec.~\ref{subsecR:governing}. We have the first law of Eq.~(\ref{firstlaw}), the equation of motion of Eq.~(\ref{Rdotdot}), the relation $m = 4\pi R^2 \sigma$, and the equation of state $p = p(\sigma)$. 

An equilibrium configuration is characterized by a compactness parameter $C$ in the way described in Sec.~\ref{subsecR:equilibria}. Denoting the equilibrium values with a zero-subscript, we have 
\begin{equation} 
m_0 = C R_0, \qquad 
\sigma_0 = \frac{C}{4\pi R_0}, \qquad 
p_0 = \frac{C^2}{16\pi R_0(1-C)}. 
\end{equation} 
The equilibrium is perturbed by letting $R$, $m$, $\sigma$, and $p$ deviate from their equilibrium values. We write $R = R_0 + \delta R$, $m = m_0 + \delta m$, $\sigma = \sigma_0 + \delta \sigma$, and $p = p_0 + \delta p$, in which all deviations (such as $\delta R$) depend on proper time $\tau$. These relations are inserted within the equations of motion, and all equations are linearized with respect to the deviations. Equation (\ref{firstlaw}) implies that 
\begin{equation} 
\delta m = -8\pi p_0 R_0\, \delta R = -\frac{C^2}{2(1-C)}\, \delta R, 
\end{equation} 
and the definition of the material mass implies that 
\begin{equation} 
\delta \sigma = \frac{1}{4\pi R_0^2}\, \delta m 
- \frac{m_0}{2\pi R_0^3}\, \delta R 
= -\frac{1}{8\pi R_0^2} \frac{C(4-3C)}{1-C}\, \delta R. 
\end{equation} 
The equation of state implies that $\delta p$ can be related to $\delta \sigma$. As previously we write the relation as $\delta p/p_0 = \Gamma\, \delta\sigma/\sigma_0$, in terms of an adiabatic exponent $\Gamma(\sigma_0)$. As was pointed out in Sec.~\ref{subsecN:stability}, $\Gamma$ should in principle be distinguished from $\gamma$ introduced previously in the context of equilibrium sequences. The exponents, however, are numerically equal when evaluated at the same reference density. 

Equation (\ref{Rdotdot}) can easily be linearized with respect to the deviations $\delta R$, $\delta m$, $\delta \sigma$, and $\delta p$, and the foregoing results imply that all deviations can be related to $\delta R$, the primary variable. After some algebra and simplification, we arrive at the perturbation equation 
\begin{equation} 
\delta \ddot{R} = -\frac{1}{4R_0^2} \frac{C}{(1-C)^2} 
\bigl[ 2(2\Gamma-3) - (7\Gamma-8)C + 3(\Gamma-1)C^2 \bigr]\, \delta R, 
\label{perturbation_equation} 
\end{equation} 
where we recall that $\Gamma$ depends on $C$ through the equilibrium density $\sigma_0$. This is again the equation for a simple harmonic oscillator, and (minus) the coefficient in front of $\delta R$ on the right-hand side is identified with the square of the oscillator's natural frequency. This reveals immediately that the equilibrium is stable when the quantity within square brackets is positive. We infer that
\begin{equation} 
\mbox{dynamical stability} \qquad \Longrightarrow \qquad 
\Gamma(C) > \Gamma_{\rm low} :=\frac{6-8C+3C^2}{(1-C)(4-3C)}. 
\label{G1lo} 
\end{equation} 
In view of Eq.~(\ref{Gamma_lo}) and the equality $\Gamma = \gamma$ for any barotropic equation of state, we see that the condition for dynamical stability is identical to the condition for $dM_0/d\sigma_0 > 0$. In other words,  the equilibrium sequence is dynamically stable so long as the gravitational mass $M_0$ increases with density, and the onset of instability occurs when $M_0$ reaches a maximum. This is the main conclusion of this section. As was pointed out in Sec.~\ref{subsecN:stability}, this conclusion rests on the assumed barotropic form for the equation of state.

\subsection{Numerical integration of the equations of motion} 
\label{subsecR:numerical} 

The stability criterion of Eq.~(\ref{G1lo}) was obtained on the basis of a perturbative analysis of the shell's equations of motion, assuming that $\delta R$ remains small at all times. The perturbation, however, does not stay small when the equilibrium is unstable. To find out what becomes of an unstable configuration, it is necessary to abandon the assumption that $\delta R$ is small, and to integrate the exact equations of motion numerically. We conclude this section with a description of methods to do this, and with a small sampling of results from numerical simulations.  

The shell's equations of motion are given by Eqs.~(\ref{firstlaw}) and (\ref{Rdotdot}). For a numerical treatment
we convert them to the the first-order system 
\begin{subequations} 
\begin{align} 
\dot{R} &= V, \\ 
\dot{V} &= -\sqrt{1+V^2} \biggl[ \frac{m}{2R^2} 
- \frac{8\pi p R}{m} \Bigl( \sqrt{1+V^2} - m/R \Bigr) \biggr], \\ 
\dot{m} &= -8\pi p R V, 
\end{align} 
\end{subequations} 
which is completed with the equation of state $p = p(\sigma)$, with $\sigma = m/(4\pi R^2)$. The accuracy of the numerical integration can be monitored by computing 
\begin{equation} 
M = m\sqrt{1+V^2} - \frac{m^2}{2R} 
\end{equation} 
and verifying that it is a constant of the shell's motion.  

The integration requires initial conditions, and we choose the shell's initial state to be a perturbed equilibrium with 
\begin{equation}  
R = R_0 = \frac{C}{4\pi \sigma_0}, \qquad 
V = \epsilon, \qquad 
m = m_0 = \frac{C^2}{4\pi \sigma_0}, 
\end{equation} 
where $C := m_0/R_0$ is the compactness parameter (which can be set to any value), $\sigma_0$ is the density of the equilibrium configuration, and $\epsilon$ is a small (positive or negative) number. 

For concreteness we adopt a polytropic equation of state $p = K \sigma^{\gamma}$, where $K$ and $\gamma$ are constants. The constant $K$ can be calibrated with the help of Eq.~(\ref{p_over_sigma}), written as $p_0/\sigma_0 = C/[4(1-C)]$, where $p_0$ is the pressure of the equilibrium configuration. We have that $K = C/[4(1-C)\sigma_0^{\gamma-1}]$, and it follows that the pressure can be expressed as 
\begin{equation} 
8\pi p = 4\pi \sigma_0 \frac{C}{2(1-C)} \biggl( \frac{1}{4\pi\sigma_0} \frac{m}{R^2} \biggr)^\gamma. 
\end{equation} 
It is convenient to work in units in which $4\pi \sigma_0 \equiv 1$. Thus, mass, radius, and time shall be measured in units of $(4\pi \sigma_0)^{-1}$; $V$ is a dimensionless quantity.  

\begin{figure} 
\includegraphics[width=0.7\linewidth]{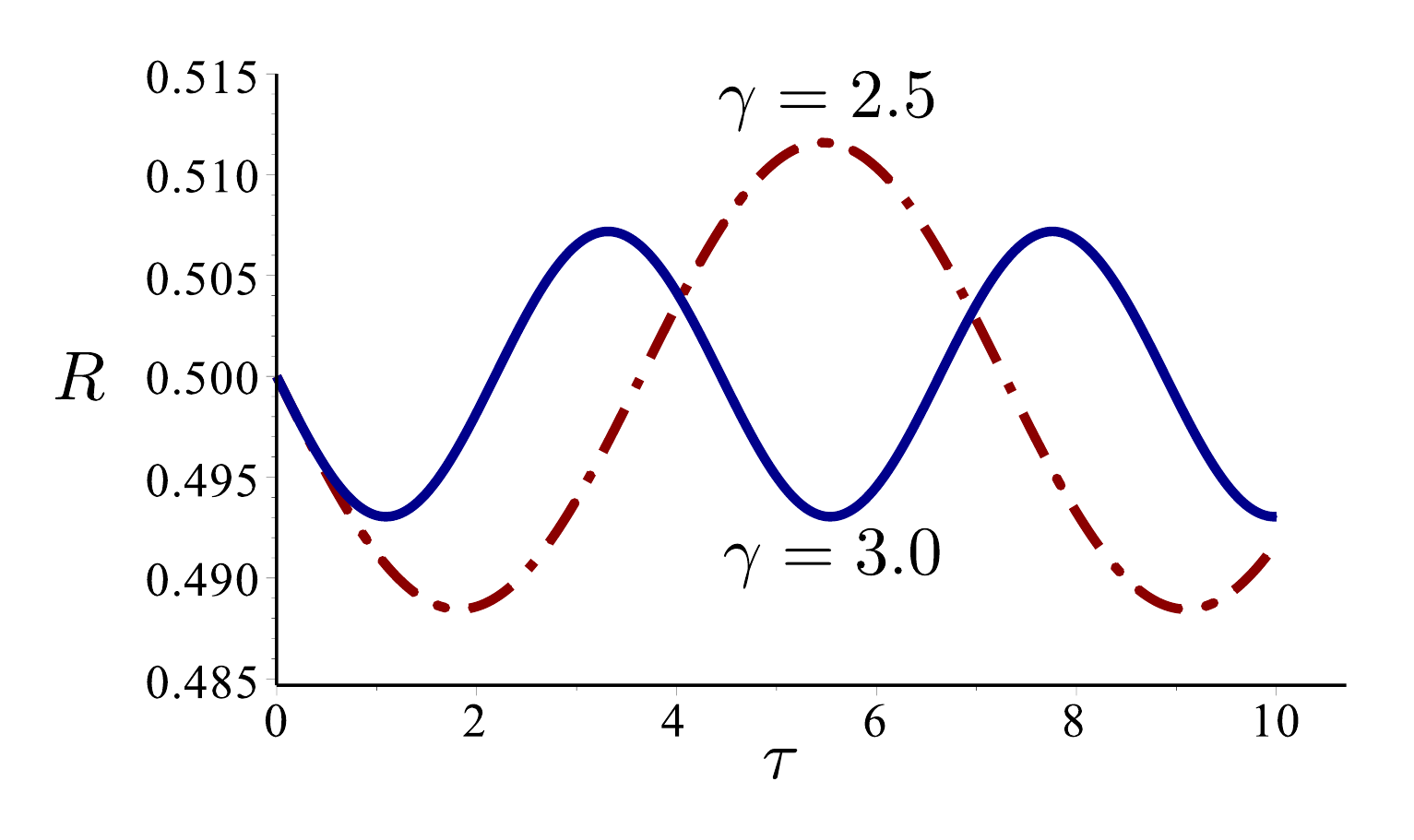}
\caption{Plot of $R$ as a function of proper time $\tau$ for two perturbed stable equilibria. Radius and time are measured in units of $(4\pi \sigma_0)^{-1}$, where $\sigma_0$ is the density of the equilibrium configuration. Both simulations adopt $C = 1/2$ and $\epsilon = -0.01$ for the initial conditions.  The dotted-dashed red curve shows a simulation with $\gamma = 2.5$, while the solid blue curve selects $\gamma = 3$. For this value of $C$, we have that in both cases $\gamma \leq \gamma_{\rm high} = 3$ and $\gamma > \gamma_{\rm low} = 11/5 \simeq 2.2$. We observe that the frequency of oscillations depends on $\gamma$, in the way described by Eq.~(\ref{perturbation_equation}).}  
\label{fig:fig5} 
\end{figure}

\begin{figure} 
\includegraphics[width=0.7\linewidth]{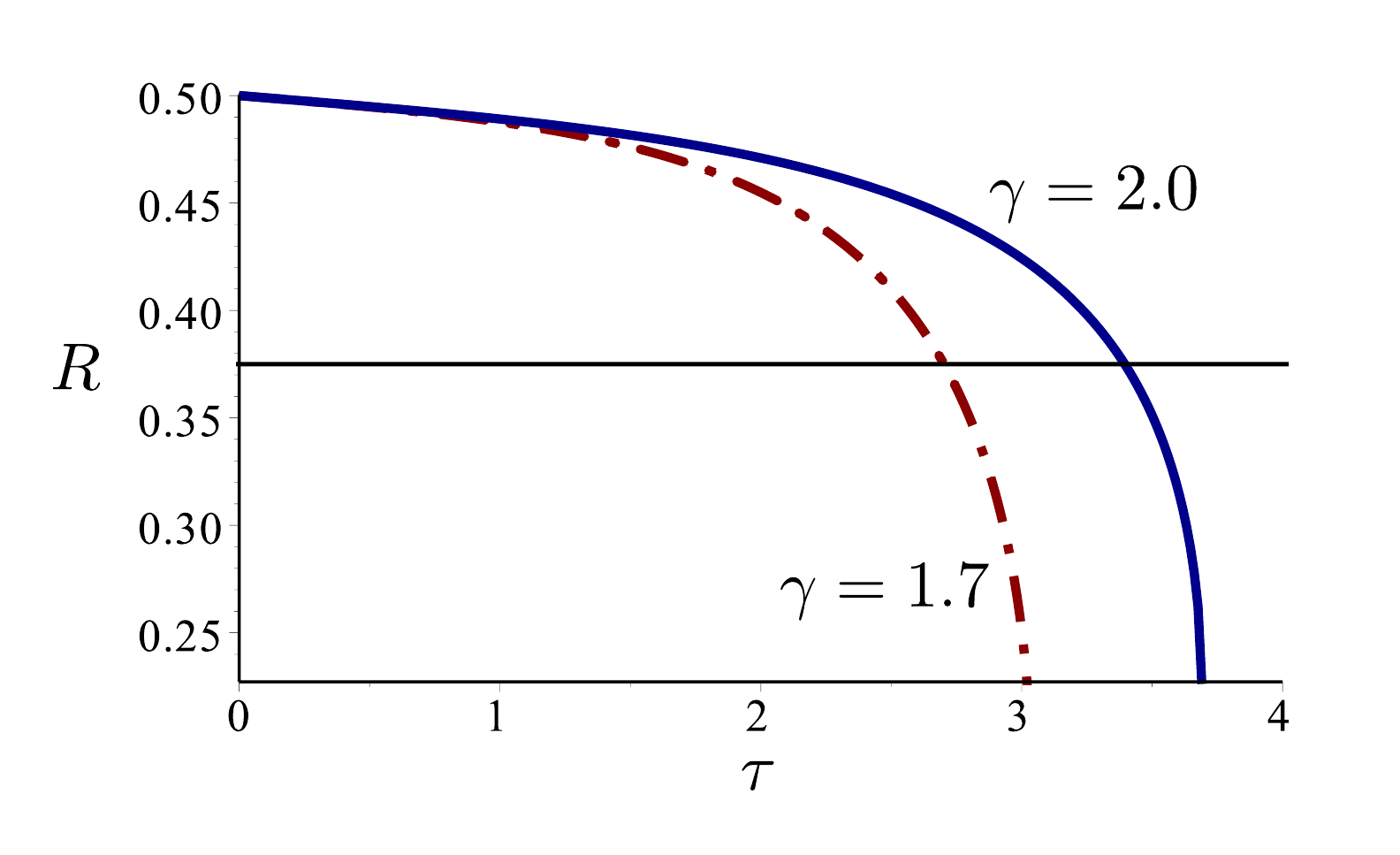}
\caption{Plot of $R$ as a function of proper time $\tau$ for two perturbed unstable equilibria. Radius, mass, and time are measured in units of $(4\pi \sigma_0)^{-1}$, where $\sigma_0$ is the density of the equilibrium configuration. Both simulations adopt $C = 1/2$ and $\epsilon = -0.01$ for the initial conditions. For these configurations we have that $M \simeq C^2(1-\frac{1}{2} C) = 3/16 \simeq 0.1875$, and the black hole forms when $R = 2M \simeq 0.3750$; this is indicated by the black horizontal line. The dotted-dashed red curve shows a simulation with $\gamma = 1.7$, while the solid blue curve selects $\gamma = 2$. For this value of $C$, we have that in both cases $\gamma < \gamma_{\rm high} = 3$ and $\gamma < \gamma_{\rm low} = 11/5 = 2.2$. The shell with $\gamma = 1.7$ provides less pressure than the shell with $\gamma = 2$, and it therefore collapses faster to a spacetime singularity at $R = 0$.} 
\label{fig:fig6} 
\end{figure}

The integration requires the selection of $\gamma$ for the equation of state, and values for $C$ and $\epsilon$ for the initial conditions. In all our simulations we ensure that $\gamma < \gamma_{\rm high} = (2-C)/(1-C)$. In Fig.~\ref{fig:fig5} we observe that the numerical integration produces oscillations in $R$ when the equilibrium configuration is stable according to Eq.~(\ref{G1lo}); in this situation $\delta R$ does remain small, and its behavior is well described by Eq.~(\ref{perturbation_equation}). When, however, we choose $\gamma$ such that  
\begin{equation} 
\gamma < \gamma_{\rm low} = \frac{6-8C+3C^2}{(1-C)(4-3C)}, 
\end{equation} 
the equilibrium is unstable. As we see in Fig.~\ref{fig:fig6}, the numerical integration describes a gravitational collapse to a black hole when $\epsilon < 0$; it would describe an unbounded expansion of the shell when $\epsilon > 0$. 

\section{Conclusion}
\label{sec:conclusion}

We have examined the statics and dynamics of thin spherical shells in both Newtonian gravity and general relativity. The equations that govern the behavior of self-gravitating thin shells are considerably simpler than those implicated in the description of three-dimensional fluid bodies, and as we have seen, they deliver their message in a very direct and vivid manner. Our main goal was to reveal the link between the maximum mass on a sequence of equilibrium configurations and the onset of dynamical instability. While a demonstration of this link requires a lengthy and difficult analysis in the case of fluid bodies, it is revealed with very little infrastructure in the case of thin  shells.   

In the case of a Newtonian shell, we showed that the maximum mass and the onset of instability occur when the effective polytropic exponent $\gamma$ decreases through the value $3/2$. This is different from what would be obtained for a realistic stellar model involving a three-dimensional distribution of fluid. In this case \cite{shapiro-teukolsky:83} the condition reads $\langle \gamma \rangle = 4/3$, where $\langle \gamma \rangle := \int \gamma p\, dV/\int p\, dV$ is the exponent averaged over the fluid's pressure profile ($dV$ is the volume element). To be sure, the results are quantitatively different. But what is important is that they are qualitatively similar: in both cases the dynamical instability is triggered when $\gamma$ drops below a certain critical value, when the body fails to provide a sufficient $dp$ for a given change of density. 

In the case of a relativistic shell, we showed that maximum mass and onset of instability occur when
\begin{equation}
\gamma = \frac{6-8C+3C^2}{(1-C)(4-3C)},
\label{tmp} 
\end{equation}
where $C = Gm/(Rc^2)$ is the shell's compactness. When $C$ is small, this becomes $\gamma \simeq 3/2 + 5C/8$. From this we see that the Newtonian limit of $3/2$ is recovered when the shell radius $R$ is very large compared with $Gm/c^2$. More importantly, we see that in the relativistic context, $\gamma$ must be larger than $3/2$ in order to ensure stability; relativistic gravity is stronger than Newtonian gravity, and a stable shell requires a larger $dp$ for the same change in density. This conclusion is not limited to small values of $C$; the function on the right-hand side of Eq.~(\ref{tmp}) is larger than $3/2$ for all values of $C$ in the interval $0 < C < 1$.    

The observation that stability requires a larger $dp$ in relativistic gravity than in Newtonian gravity is true also in the case of a three-dimensional distribution of fluid. In this case \cite{shapiro-teukolsky:83, chandrasekhar:64b} it is not possible to write the criterion as in Eq.~(\ref{tmp}), with a simple, universal function of $C$ on the right-hand side. For small $C$, however, the criterion becomes the simple $\langle\gamma\rangle = 4/3 + kC + O(C^2)$, where $\langle\gamma\rangle$ is again the pressure-averaged polytropic exponent, and $k > 0$ is a numerical factor of order unity that depends on the details of internal structure. Here also the main message is that $\langle\gamma\rangle$ must be {\it larger than} its Newtonian value.    

We have emphasized that the link between maximum mass and onset of dynamical instability relies on the  assumption that the equation of state is of the barotropic form $p = p(\sigma)$. Equations of state implicating additional variables would not produce the crucial equality between the effective polytropic index $\gamma$ (which is of relevance to equilibrium sequences) and the adiabatic index $\Gamma$ (which is of relevance to the stability analysis); as a result, the maximum mass and the onset of instability would occur at different points on the equilibrium sequence. This restriction applies equally well to three-dimensional fluid bodies: here also the link between maximum mass and onset of instability relies on a barotropic equation of state.  

Throughout the paper we have thought of a thin shell as a crude proxy for a realistic neutron star. Our results, however, apply independently of this association, and indeed, our thin shell could be used as a substitute for any self-gravitating body with a barotropic equation of state. This object, for example, could be a white dwarf modeled as a free, relativistic, and degenerate electron gas. That we have kept our sight on neutron stars merely reflects a personal bias in favor of relativistic gravity, which is required for an accurate description of neutron stars and black holes, but not for white dwarfs.  

\begin{acknowledgments} 
This work was supported by the Natural Sciences and Engineering Research Council of Canada.   
\end{acknowledgments} 

\appendix

\section{Regularization of the gravitational force} 
\label{sec:A} 

In this Appendix we justify the expression $Gm^2/(2R^2)$ used in Sec.~\ref{subsecN:governing} for the gravitational force acting on a thin spherical shell of radius $R$, density $\sigma$, and mass $m$. We consider a thick spherical shell, described by a smooth volume density $\rho(r)$ that is assumed to be sharply peaked around $r = R$. We let 
\begin{equation} 
\mu(r) = 4\pi \int_0^r \rho(r') r^{\prime 2}\, dr' 
\end{equation} 
be the mass inside radius $r$; this is zero inside the shell, equal to $m$ outside, and the mass function rises sharply from zero to $m$ in the vicinity of $r = R$. Because $\rho = \mu'/(4\pi r^2)$, with a prime indicating differentiation with respect to $r$, we have that $\mu'$ is sharply peaked around $r=R$. 

We calculate the total gravitational force on the thick shell. The force acting on a layer of mass $d\mu$ situated at radius $r$ is $G \mu\, d\mu/r^2$, and integrating across the entire shell gives  
\begin{equation} 
F = G \int_0^\infty \frac{\mu \mu'}{r^2}\, dr 
\end{equation} 
for the total force. Because $\mu'$ is sharply peaked around $r = R$, the force can be well approximated by 
\begin{equation} 
F = \frac{G}{R^2} \int_0^\infty \mu \mu'\, dr 
= \frac{G}{2 R^2} \int_0^{m^2} d\mu^2 
= \frac{Gm^2}{2 R^2}. 
\end{equation} 
This expression remains well defined in the limit of an infinitely thin shell, and it is precisely the expression used in Sec.~\ref{subsecN:governing}. 

\section{Relativistic thin shells} 
\label{sec:B}

The equations that govern the motion of a thin spherical shell in general relativity are derived in Sec.~3.9 of Ref.~\cite{poisson:b04}, building on Israel's surface-layer formalism \cite{israel:66}. We import these results here, and  bring them to the form displayed in Sec.~\ref{subsecR:governing}.  

The governing equations are  
\begin{equation} 
\sigma = -\frac{\beta_+ - \beta_-}{4\pi R}, \qquad
\sigma + 2p = \frac{\dot{\beta}_+ - \dot{\beta}_-}{4\pi \dot{R}},
\label{geq1} 
\end{equation} 
where $R(\tau)$ is the shell's radius expressed in terms of the proper time $\tau$ of an observer comoving with the shell, $\sigma(\tau)$ is the shell's mass density (mass per unit area), and $p(\tau)$ is the shell's tangential pressure. An overdot indicates differentiation with respect to $\tau$, and we have introduced 
\begin{equation} 
\beta_+ := \bigl( 1 + \dot{R}^2 - 2M/R \bigr)^{1/2}, \qquad 
\beta_- := \bigl( 1 + \dot{R}^2 \bigr)^{1/2}, 
\label{geq2} 
\end{equation} 
where $M$ is the shell's gravitational mass --- the mass parameter of the Schwarzschild spacetime outside the shell. For our manipulations it is useful to note that 
\begin{equation} 
\dot{\beta}_+ = \frac{\ddot{R} + M/R^2}{\beta_+}\, \dot{R}, \qquad 
\dot{\beta}_- = \frac{\ddot{R}}{\beta_-}\, \dot{R}. 
\label{beta_dot} 
\end{equation} 
The shell's material mass $m$ is defined by $m := 4\pi R^2 \sigma$, and this is to be carefully distinguished from the gravitational mass $M$. The shell's area is $A := 4\pi R^2$.  

Equations (\ref{geq1}) imply that $(\sigma + 2p) dR/d\tau = -d(\sigma R)/d\tau$, which can be put in the form 
\begin{equation} 
\frac{d}{d\tau} (\sigma R^2) + p \frac{d}{d\tau} R^2 = 0. 
\end{equation} 
This is recognized as $\dot{m} = -p\, \dot{A}$, the first law of thermodynamics, stated in Eq.~(\ref{firstlaw}).  

Taking into account the relation $m = 4\pi R^2 \sigma$, the first of Eqs.~(\ref{geq1}) can be written as 
\begin{equation} 
\beta_+ - \beta_- = -\frac{m}{R}. 
\label{fi_1} 
\end{equation} 
Multiplying both sides by $\beta_+ + \beta_-$ and incorporating Eqs.~(\ref{geq2}), we get
\begin{equation} 
\beta_+ + \beta_- = \frac{2M}{m} 
\label{fi_2}
\end{equation}  
after simple manipulations. Adding and subtracting the preceding equations, we have that 
\begin{equation} 
m \beta_+ = M - \frac{m^2}{2R}, \qquad 
m \beta_- = M + \frac{m^2}{2R}. 
\end{equation} 
The second equation is the same as Eq.~(\ref{M_eq}). A second-order differential equation for $R(\tau)$ is obtained by a straightforward differentiation of Eq.~(\ref{M_eq}); this produces Eq.~(\ref{Rdotdot}).  

%\bibliography{../bib/master}
\bibliography{shell} 

\end{document}